\documentclass[aps,prl,twocolumn,groupedaddress,showpacs]{revtex4}

\usepackage{graphicx}

\begin{document}
    \title{Closed-loop Quantum Parameter Estimation:
        Spins in a Magnetic Field}
    \author{JM Geremia}
    \email{jgeremia@Caltech.EDU}
    \author{John K. Stockton}
    \author{Hideo Mabuchi}
    \affiliation{Norman Bridge Laboratory of Physics,
    California Institute of Technology, Pasadena CA 91125}

\begin{abstract}
We present an experimental demonstration of closed-loop quantum
parameter estimation in which real-time feedback is used to
achieve robustness to modeling uncertainty. By performing
broadband estimation of a magnetic field acting on hyperfine spins
in a cold atom ensemble, we show that accuracy is not compromised
by fluctuations in total atom number even though the measured
signal in our canonical configuration depends only on the product
of the field and atom number. This methodology could be essential
for efforts to utilize conditional squeezing in spin-resonance
measurements.
\end{abstract}


\maketitle

\noindent Optimal design of experimental procedures and data
analysis strategies can often be accomplished using techniques
from parameter estimation theory. Such an approach can be
essential for obtaining acceptable performance in scenarios where
modeling is subject to some degree of uncertainty, and has seen
widespread use in fields that lie near the interface of physics
and information science. Prominent current examples of such fields
include metrology, optical communication, and computation with
novel substrates. As micro- and nano-scale systems with manifestly
non-classical behavior have gained importance in these areas,
researchers have devoted increasing attention to the extension of
parameter estimation methodologies to problems involving quantum
mechanical inference and dynamics. Our aim in this article is to
establish that \textit{real-time feedback} plays a central role in
the quantum regime--- as it does in classical scenarios---
enabling robust parameter estimation in the presence of
significant modeling uncertainty.

\begin{figure}[b]
    \vspace{-5mm}
    \begin{center}
    \includegraphics{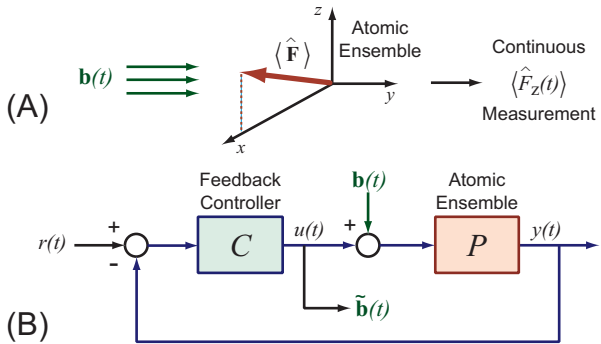}
    \end{center}
    \vspace{-3mm} \caption{Quantum parameter
    estimation of a magnetic field by observing
    ensemble spin dynamics according to both an
    open (A) and closed (B) loop experimental methodology.
    \label{Figure::MeasurementSchematic}}
\end{figure}

In quantum parameter estimation the central objective is to
extract information about a static or time-dependent parameter in
a Hamiltonian, via direct or indirect measurements performed on a
system that evolves according to this Hamiltonian
\cite{holevo2001a,Gambetta2001,Verstraete2001}. An elementary
example of such a process is estimating the amplitude of a
magnetic field, $\mathbf{b}(t)$, by observing short-time Larmor
precession of a spin ensemble \cite{Dupont1969} (here we will
consider an ensemble consisting of hyperfine spins in a cloud of
laser-cooled atoms). The field-spin interaction is described by a
magnetic dipole Hamiltonian,
\begin{equation}
  \hat{H}[t;\mathbf{b}(t)] = \hbar \gamma \mathbf{b}(t) \cdot
    \hat\mathbf{F}
\end{equation}
where $\gamma$ is the gyromagnetic ratio and $\hat\mathbf{F}$ is
the total angular momentum operator for the ensemble.

A canonical estimation procedure is depicted in Fig.\
\ref{Figure::MeasurementSchematic}(A).  An atomic spin ensemble is
prepared such that its net magnetization, or Bloch vector,
$\langle \hat\mathbf{F} \rangle$, achieves a coherent spin state
along the $x$-axis. In the presence of an external magnetic field,
$\mathbf{b}(t)$, the bulk magnetization precesses around
$\mathbf{b}$ with instantaneous frequency, $\omega_\mathrm{L}(t) =
\gamma |\mathbf{b}(t)|$. Generally, it is arranged such that
$\mathbf{b}(t)=b(t)\hat\mathbf{y}$ lies along the $y$-axis so that
the spin state can be resolved from a continuous measurement of
$\langle\hat{F}_\mathrm{z}\rangle$ \cite{Jessen2003,Deutsch2003}.
An estimate, $\tilde\mathbf{b}$, of the external magnetic field is
then obtained (by regression or filtering) from the small-angle
relation
\begin{equation} \label{Equation::SimpleEstimator}
    \tilde\mathbf{b} = \frac{\langle\hat{F}_\mathrm{z}(t)\rangle}
        {\gamma |\hat\mathbf{F}(t)| t} \, \hat\mathbf{y}
        , \quad t \ll \omega_\mathrm{L}^{-1}.
\end{equation}
This procedure can achieve Heisenberg limited sensitivity
\cite{Geremia2003} by exploiting conditional spin-squeezing
\cite{Kitagawa1993,Kuzmich2000}. The problem with this approach is
that it requires accurate knowledge of the net magnetization,
$|\hat\mathbf{F}(t)|$, a quantity that unfortunately varies due to
decoherence and shot-to-shot fluctuations in the atom number.
Uncertainty in $|\hat\mathbf{F}(t)|$ directly translates into
uncertainty in $\tilde\mathbf{b}(t)$.

Fig.\ \ref{Figure::MeasurementSchematic}(B) depicts an alternative
estimation procedure that is robust to fluctuations in
$|\hat\mathbf{F}(t)|$ \cite{Thomsen2002}. The spin ensemble and
$\langle \hat{F}_\mathrm{z}\rangle$ measurement are situated
within a feedback control loop that attempts to stabilize $\langle
\hat{F}_\mathrm{z}\rangle$ to a reference value, $r(t)=0$. In the
presence of a time-varying magnetic field signal, $\mathbf{b}(t)=
b(t)\hat\mathbf{y}$, the controller imposes a compensating field,
$\mathbf{b}_\mathrm{c}(t) \simeq -b(t) \hat\mathbf{y}$, to try to
suppress the atomic Larmor precession. The closed-loop estimate is
then given by $\tilde{\mathbf{b}}(t)=-\mathbf{b}_\mathrm{c}$, and
its accuracy is determined by the controller's ability to respond
promptly and accurately to changes in $\mathbf{b}(t)$. In this
work we demonstrate that standard design techniques enable the
implementation of feedback controllers with excellent tracking and
high robustness to fluctuations in $|\hat\mathbf{F}(t)|$. This
illustrates new utility for real-time feedback in cold atom
physics, as previous investigations have focused on active control
of atomic motion \cite{Raithel2002,Rempe2002}.

\begin{figure}[t]
\begin{center}
\includegraphics{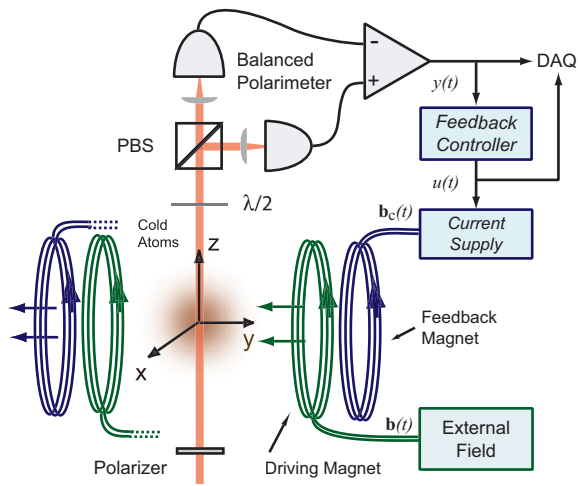}
\end{center}
\vspace{-3mm} \caption{Schematic of our apparatus for closed-loop
parameter estimation using a cold atom ensemble to determine
$\mathbf{b}(t)$ in a magnetic dipole Hamiltonian, $\hat{H}(t) =
\hbar \gamma \mathbf{b}(t) \cdot \hat\mathbf{F}$.
\label{Figure::Schematic} } \vspace{-1mm}
\end{figure}

Fig.\ \ref{Figure::Schematic} provides a schematic of our
experimental apparatus implemented according to the design in
Fig.\ \ref{Figure::MeasurementSchematic}(B). It consists of a cold
atom ensemble, a Faraday polarimeter for continuously probing the
atomic Bloch vector, and a high-speed Helmholtz coil along the
$y$-axis for applying feedback.  The spin system is provided by
the 6$^2$S$_{1/2}$(F=4) ground state hyperfine manifold in Cs,
which contains $(2F+1)=9$ energetically degenerate Zeeman
sublevels in zero field. Therefore, the net angular momentum of
the polarized spin state is given by, $N \hbar
\sqrt{F(F+1)}\approx 4 N\hbar$, for an ensemble of $N$ atoms.

Cold atom samples were obtained by loading $N\sim 10^9$ neutral
$^{133}$Cs atoms into a magneto-optical trap (MOT) from a $\sim 1
\times 10^{-8}$ Torr background vapor, using optical trapping
beams (30 mW each, 2.5 cm diameter) derived from an
injection-locked diode laser. The atoms were cooled to 1 $\mu$K
via a 5 ms $\sigma_+$/$\sigma_-$ polarization gradient cooling
phase, and a coherent spin state was produced by optical pumping
with $\sigma_+$ polarized light (100 $\mu$W with a 6.2 mm Gaussian
waist) tuned to the 6$^2$S$_{1/2}$(F=4) $\rightarrow$6$^2
$P$_{3/2}$(F$^\prime$=4) hyperfine transition. A 45 mW, 2.5 cm
diameter re-pumping beam was used.

Continuous weak measurement of $\hat{F}_\mathrm{z}$ was
implemented using a free-running diode laser, blue-detuned from
the (F=4)$\rightarrow$(F$^\prime$=5) transition by $\Delta$=2 GHz.
The 65 $\mu$W probe beam was linearly polarized by a high
extinction ($>$10$^6$) Glan-Thompson prism polarizer and
mode-matched to the Gaussian waist of the atomic spatial
distribution. Relaxation of the coherent spin state due to the
probe light was measured to be $T_2$ = 11.2 ms.

Faraday rotation of the probe light was detected using a balanced
polarimeter (1 MHz detector bandwidth) constructed from a high
extinction Glan-Thompson polarizing beam splitter.  This
configuration produces a photocurrent, $y(t)$, proportional to the
$z$-component of the spin angular momentum
\cite{Thomsen2002,Deutsch2003,Jessen2003},
\begin{equation} \label{Equation::Photocurrent}
    y(t) = 2 \sqrt{M} \langle \hat{F}_\mathrm{z}(t)\rangle + \zeta(t)
\end{equation}
where $M$ is the measurement strength and $\zeta(t)$ reflects
measurement noise. Background magnetic fields were nulled to
$<$100 $\mu$G with large (d = 1 m) external three-axis Helmholtz
coils by Larmor precession measurements.

The feedback system in Fig.\ \ref{Figure::Schematic} utilizes the
photocurrent, $y(t)$, to control the strength of a Hamiltonian,
\begin{equation}
    \hat{H}_\mathrm{c}(t)  =  \hbar \gamma b_\mathrm{c}[y(t)] \,
    \hat{F}_y
\end{equation}
that rotates the Bloch vector around the $y$-axis.  It is the job
of the \textit{controller} to determine the appropriate feedback
strength, $b_\mathrm{c}(t) = \beta u(t)$, based on the observation
$y(t)$. For a linear controller,
\begin{equation}
    u(t)  =  \int_0^t C(t-\tau) y(\tau)\, d\tau ,
\end{equation}
\vspace{-.5mm}where $C(t)$ must be designed to satisfy tracking
and robustness objectives. The controller output, $u(t)$, is used
to program a current supply that drives a $y$-axis Helmholtz coil
surrounding the atomic sample.  This requires accurate calibration
(obtained in our case from Larmor frequency measurements) of the
gain, $\beta$, from current supply programming voltage, $u(t)$, to
the feedback field, $b_\mathrm{c}(t)$.

\begin{figure*}
\vspace{-2mm}
\begin{center}
\includegraphics{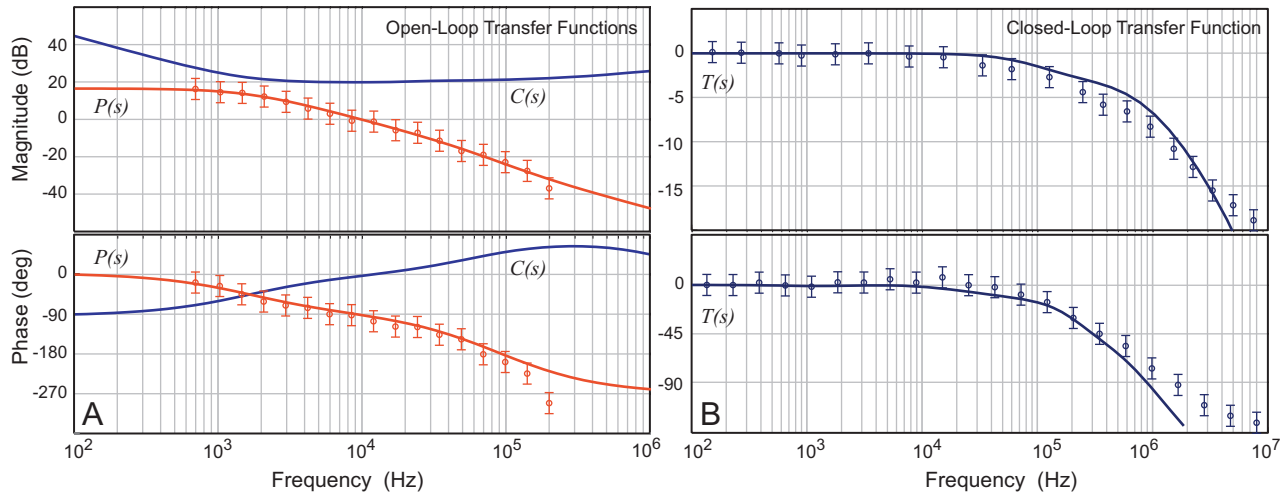}
\end{center}\vspace{-6mm}
\caption{(A) Measured (points) and fitted (solid line) plant
transfer function, $P(s)$, between the feedback signal, $u(t)$,
and polarimeter photocurrent, $y(t)$, as well as the designed
controller, $C(s)$.  (B) Measured (points) and calculated (solid
line) closed-loop transfer function, $T(s)$ which shows a feedback
bandwidth of 100 kHz.\label{Figure::BodePlots} }
\end{figure*}

In control theory, it is customary to design $C(t)$ in the
frequency rather than time domain by taking Laplace transforms of
the functions, $C(t)\rightarrow C(s)$, $y(t)\rightarrow y(s)$,
etc., where $s=j\omega$ and $j=\sqrt{-1}$.  This results in the
closed-loop frequency response of the control system,
\vspace{-.5mm}
\begin{equation}
    T(s) = \frac{y(s)}{r(s)}=\frac{C(s)P(s)}{1+C(s)P(s)}
\end{equation}
\vspace{-.5mm}where $P(s)=y(s)/u(s)$ is the transfer function from
the feedback coil programming voltage to the photocurrent. The
tracking objective is to adjust the gain and phase of $C(s)$ such
that $T(s)\simeq 1$ over as large a frequency range as possible.
Standard results from robust loop-shaping theory
\cite{Doyle1990,Doyle1997} lead to a controller design,
\vspace{-.5mm}
\begin{equation} \label{Equation::Controller}
  C(s) = \frac{Q(s)}{1+P(s)Q(s)},
  \quad Q = \frac{W(s)}{P_{\mathrm{mp}}(s)}
\end{equation}
\vspace{-.5mm}where $W(s)$ is a stable, strictly proper
[$W(s\rightarrow \infty)=0$] function that adjusts the bandwidth
of $C(s)$ and $P_{\mathrm{mp}}(s)$ is the minimum phase
contribution to $P(s)$.

Ideally, the Helmholtz coil current supply would introduce no
additional frequency dependence to $P(s)$, so that $P(s)$ would be
determined solely by atomic dynamics.  In the limit, $\langle
\hat{F}_\mathrm{z}\rangle \ll | \hat\mathbf{F} |$, which is
maintained in closed-loop, this would be [using $\langle
\hat{F}_\mathrm{z} (t)\rangle = \exp(-t/T_2)\sin(\omega_L t)$],
\begin{equation}
    P(s) \approx 2\gamma\sqrt{M}|\hat\mathbf{F}|\frac{1}{s}
\end{equation}
and proportional feedback [constant $C(s)$] would provide good
tracking.  However, it is not always possible to construct a
current supply with sufficient bandwidth. Finite available supply
power places an upper bound on the closed-loop bandwidth for
driving an inductive load, and a more intelligent $C(s)$ design is
necessary. Our Helmholtz coil supplies were constructed using
high-power (500 W) operational amplifiers and displayed 73 degrees
of phase delay at 100 kHz. With proportional control, this would
be insufficient phase margin.

Fig.\ \ref{Figure::BodePlots} shows the measured transfer
function, $P(s)$, for the combined atomic ensemble, Faraday
polarimeter and feedback coil system.  It was generated by
utilizing a network analyzer to perform a swept sine analysis of
$P(s)$.  For each data point, the analyzer was triggered following
preparation of a coherent spin state according to the cooling and
optical pumping procedure described above.  Although $P(s)$ is
approximately an integrator, it displays substantially larger
phase delay at higher frequency; a fit of the transfer function
yields the model,
\begin{equation} \label{Equation::PlantTF}
    P(s) = \frac{1.6 \times 10^4(8.0 \times 10^5-s)}
    {s^2+4.1 \times 10^5 s + 4.0 \times 10^9} \,.
\end{equation}
By factoring this transfer function into its minimum-phase and
all-pass components, $P(s)=P_{\mathrm{mp}}(s) P_{\mathrm{ap}}(s)$,
a stabilizing feedback controller was obtained using Eq.
(\ref{Equation::Controller}). $W(s)$ was chosen to be a
single-pole (Butterworth) low-pass filter with $f_\mathrm{c}=1$
MHz which yielded the transfer function, $C(s)$, in Fig.\
\ref{Figure::BodePlots}(A).  The feedback controller was
implemented using high bandwidth analog electronics and resulted
in the closed-loop transfer function, $T(s)$, depicted in Fig.\
\ref{Figure::BodePlots}(B). The significance of $T(s)$ is that
$\|1-T(s)\|$ provides a measure of the tracking error, and thus
the error in the parameter estimation.

\begin{figure*}
\vspace{-3mm}
\begin{center}
\includegraphics{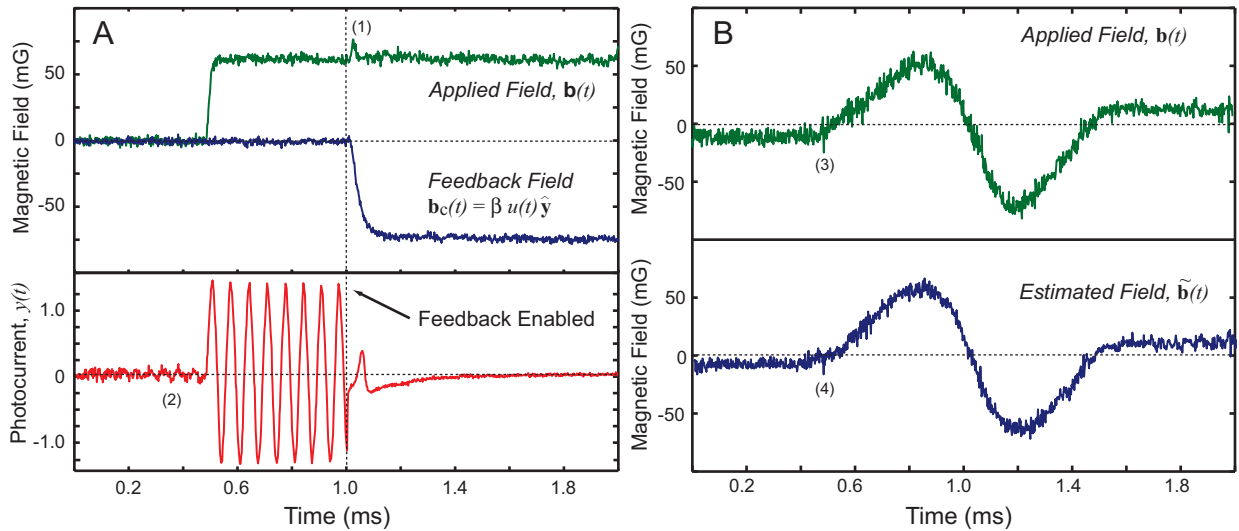}
\end{center}
\vspace{-6mm} \caption{Single-shot, closed-loop parameter
estimation of a stationary (A) and time-varying (B) applied
magnetic field $\mathbf{b}(t)$.  In (A) feedback is enabled at t =
1 ms (dotted line).  The labels identify several artifacts in the
data: (1) cross-talk between the driving and feedback coils, (2)
background field fluctuations due to power supply noise at 51.3
kHz, and (3) noise in the driving field that is also revealed by
the estimator (4). \label{Figure::Tracking} }
\end{figure*}

Fig.\ \ref{Figure::Tracking} shows a demonstration of the
closed-loop parameter estimation procedure for both stationary and
time-dependent magnetic fields generated by an auxiliary $y$-axis
Helmholtz coil (refer to Fig.\ \ref{Figure::Schematic}). Plot (A)
indicates the effect of real-time feedback on the polarimetry
photocurrent, $y(t)$.  At time, $t$ = 0.5 ms, a field of
$B_\mathrm{y}$ = 50 mG was turned on, with feedback disabled,
resulting in atomic Larmor precession. However, when the feedback
loop was closed at $t$ = 1 ms, it acted to null out the applied
field and lock the Bloch vector into the $xy$-plane.  This
simultaneously provided an estimation of the applied field,
$\tilde{\mathbf{b}} = -\mathbf{b}_\mathrm{c} = -\beta u(t)
\hat\mathbf{y}$. Plot (B) demonstrates our capability to track
time-varying fields within the feedback bandwidth of 100 kHz;
here, feedback was enabled the entire time during a 5 kHz
bandwidth applied field.

\begin{figure}[b]
\vspace{-3mm}
\begin{center}
\includegraphics{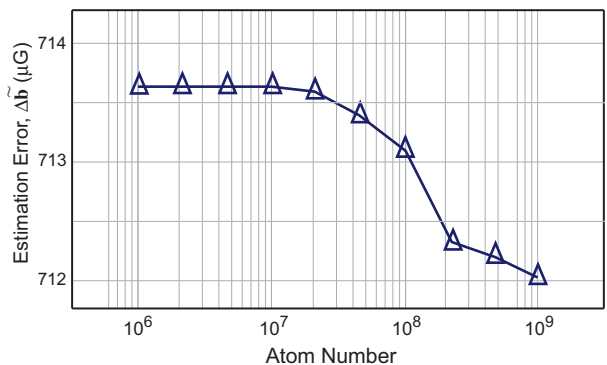}
\end{center}
\vspace{-6mm}\caption{Robustness of the single-shot, closed-loop
estimation to atom number fluctuations (refer to text).
\label{Figure::TrackingError} }
\end{figure}

To demonstrate robustness, the parameter estimation error was
measured as a function of atom number, $N$, by varying the MOT
loading time, $t_\mathrm{L}$, according to a calibration of $N$
versus $t_\mathrm{L}$, (accurate to 40\%) obtained from resonant
fluorescence imaging. The estimation error, $\Delta
\tilde\mathbf{b}$, was computed from ensemble averages,
$E[\|\mathbf{b}(t)-\tilde\mathbf{b}(t)\|_2]$, over 100 replicate
measurements for each sampled value of $N$ according to the
semi-norm,\vspace{-.5mm}
\begin{equation}
    \| \mathbf{b}(t) \|_2 = \frac{1}{T} \int_0^T
    |\mathbf{b}(t)|^2 \, dt \,.
\end{equation} \vspace{-.5mm}
As seen in Fig.\ \ref{Figure::TrackingError}, $\Delta
\tilde{\mathbf{b}}$ was essentially unaffected by fluctuations in
$N$ spanning three orders of magnitude.  The residual estimation
uncertainty of approximately 713 $\mu$G is the result of tracking
error due to finite controller gain which provides motivation for
higher bandwidth feedback systems.  A closed-loop estimator with a
1 MHz unity-gain point is expected to provide $\sim$10 nG field
resolution similar to current magnetometers \cite{Romalis2003}.

Although the current experiment did not produce an appreciable
degree of spin-squeezing, it has demonstrated a connection between
feedback and robustness (to inevitable atom-number fluctuations)
that will apply equally well in future experiments with improved
sensitivity \cite{Thomsen2002}. Hence the closed-loop methodology
we advocate should enable--- and may even be essential for--- the
implementation of proposals to utilize conditional spin-squeezing
for sensitivity beyond the Standard Quantum Limit in various
applications of spin resonance with cold atoms, such as
magnetometry \cite{Stockton2003}, atomic frequency standards, and
matter-wave gravimetry.

This work was supported by the NSF (PHY-9987541, EIA-0086038), the
ONR (N00014-00-1-0479), and the Caltech MURI Center for Quantum
Networks (DAAD19-00- 1-0374). JKS acknowledges a Hertz fellowship.
We thank Andrew Berglund and Michael Armen for experimental
assistance and Andrew Doherty and Poul Jessen for helpful
discussions. Additional information is available at
http://minty.caltech.edu/Ensemble.

\end{document}